    \documentstyle[proceedings,psfig]{crckapb} 
\def\VEV#1{\left\langle #1\right\rangle}
\def\vectheta{{\vec \theta}}

\begin{opening}
\title{Probing Density Fluctuations using the FIRST Radio Survey}
\author{C.M.Cress + FIRST collaboration}
\institute{Columbia University\\
           Astronomy Dept., 538 W120th St, New York, NY10027, USA}
\end{opening}

\runningtitle{Probing Density Fluctuations with FIRST}

\begin{document}

% The \begin{document} command comes after the \end{opening}
% command.

\section{Introduction}
For decades, the clustering of luminous objects has been
used as an indirect probe of the density fluctuations in the universe. 
Although the relationship between luminous objects and mass in the universe
is not fully understood, it has been hoped that the observed evolution of
clustering of different luminous populations might constrain many 
cosmological unknowns -- the geometry 
of the universe, the nature of any non-luminous matter, the nature of 
density fluctuations in the early universe and hence the physics of the
early universe. Indeed, models for structure formation
which are based on variants of the `cold dark matter' (CDM) model
have received much attention largely because they have been fairly
successful at predicting the clustering of luminous objects. 

Much information on the clustering of nearby populations has been 
obtained from optical and infra-red surveys of galaxies (and of clusters 
of galaxies). In addition,
Peacock and Nicholson (PN, 1991) investigated the clustering of
bright ($S_{2.7}>1$Jy) nearby ($0.01<z<0.1$) radio galaxies.
The new radio surveys provide an opportunity for investigating
the clustering of much fainter radio sources. Probing density
fluctuations with a different tracer provides interesting information
on the relative `bias' of different populations and, of course,  
one might hope that the great depth 
and large area covered by the new surveys can be used to probe 
clustering at higher redshift and on larger scales than was previously 
possible.

\section{Measuring the correlation function in the FIRST Survey}
The FIRST survey has now been extended to include a total of
$3000\rm\,deg^2$ of sky yielding a catalog of about 250,000 sources 
with $22^\circ<\delta<42^\circ$.
and $7^h30^m<\alpha<17^h30^m$. Details of the mapping and catalog generation 
can be found in Becker et al. (1996) and in White et al. (1997). 
Details of how the angular correlation function (CF) is measured can be found
in Cress et al. (1996). Here, we use the CF estimator given by Landy and 
Szalay (LS, 1993): 
$w(\theta)={[DD(\theta)-2DR(\theta)+RR(\theta)]/{RR(\theta)}}$.
While single objects are sometimes resolved into two or more sources in
the catalog, it appears that these extra sources at small separations 
do not contribute to the correlation function at larger separations
(see Cress et al. 1996). Error bars are obtained using the 
`partition bootstrap method'
in which the CF is calculated for 10 subdivisions of the survey region
and the standard deviation of these measurements at each angle is used
as a measure of the error. All areas in the survey where the rms noise
is less than 0.2 mJy were used in the analysis. CF parameters were
determined from a straight line fit to the log-log plots. 

We also investigated the CF of FIRST sources in the 
catalog that have a galaxy with $E<19$ within $3^{\prime\prime}$ in the 
APM catalog of the POSS\,I survey plates.

Writing the angular CF as $w(\theta)=A \theta^{1-\gamma}$ with $\theta$
measured in degrees, we find
$A=0.002$, $\gamma=2.1$ for the whole sample and $A=0.04$, $\gamma=1.8$
for the APM matches. 

\begin{figure}
%\vspace{6cm}  % amount of vertical space needed
\centerline{
\psfig{figure=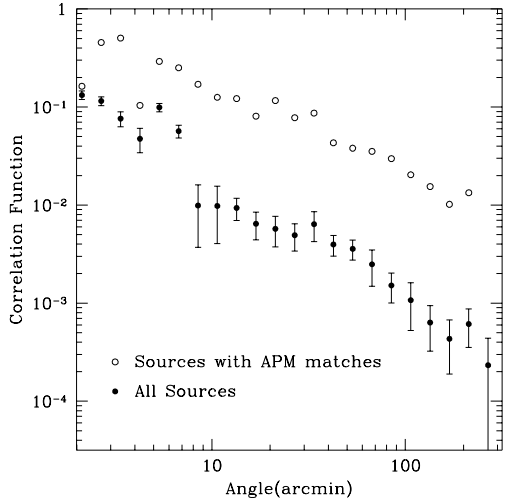,height=7cm}
}
\caption{The correlation function for (i) all sources and (ii) for those
sources which are identified with E$<$19 galaxies in the APM survey.}
\end{figure}

\section{Inferring spatial information}
The spatial CF, $\xi$, is related to the angular CF via Limber's equation. 
\begin{equation}
     w(\theta)= 2 \int_0^\infty \int_0^\infty \,
\left({dN\over dz} {1\over N} \right)^2 \xi(r,z) dx du,
\label{limbereqn}
\end{equation}
where the physical separation is given by 
$r^2=a^2(t)[u^2/F^2(x) + x^2 \theta^2]$, $u$ is the difference in radial
distance to two sources, $x$ is the average distance to the two sources,
$\theta$ is the angle between the sources in radians, $a$ is the scale 
factor and $F=1$ in a flat universe (Baugh and Efstathiou, 1993).

%\subsection{Power-law model for spatial clustering}
If $\xi$ has a power-law form and density perturbations evolve linearly 
then we can write $\xi(r,z)=(r/r_0)^{\gamma}(1+z)^{-2}$ (in comoving 
coordinates). We can then estimate $dN\over dz$ using luminosity functions
give in Dunlop and Peacock (DP, 1991) or Condon (1984). Using the LS
estimator of the CF and DP's models 1 to 5
we obtain an $r_0$ around $8h^{-1}$Mpc. DP's model 7 and Condon's 
model give slightly lower values: between $8h^{-1}$Mpc and $6h^{-1}$Mpc. 
(Note that the $r_0$ derived in Cress et al. uses the CF given by a different
estimator -- we later found the LS estimator to be more robust to changes
in survey geometry). The $r_0$ results depend
on the assumed evolution but the derived clustering appears fairly similar 
to the clustering of galaxies detected in the optical. 
  
%\subsection{CDM+inflation model for spatial clustering}
Of course, the spatial CF is not necessarily well-represented by 
a pure power law and it is interesting to take spatial CF's 
predicted by CDM models and investigate what they imply 
for the clustering of objects with the redshift distribution of
FIRST sources. CDM makes predictions for how the $mass$
clustering evolves but one must also consider the possiblity that the 
luminous sources are `biased' (Kaiser 1987) relative to the mass distribution. 

In CDM models one starts with a scale-free primordial power
spectrum, $P(k)\propto k^n$ (where $k$ is the comoving wavenumber). 
This $P(k)$ is then `processed' according a transfer function for CDM. 
At the epoch at which structure begins to grow $P(k)=P_0\,k^n\,T^2(k)$,
where $T(k)$ for CDM is given by Bond \& Efstathiou (1984) and $P_0$ is 
the normalisation. We normalise to COBE 4-year data using Liddle et al.'s
(1996) formula which is valid for spatially flat models.
 
In the linear regime, the time evolution of the power spectrum can be
calculated analytically and depends only on the expansion of the
Universe. It is given by 
$P(k,z) = P(k,z=0)\times G^2(\Omega_0(z),\Omega_\Lambda(z))/(1+z)^2 $
where the growth factor, $G$, is given in Carroll, Press \& Turner 
(1992). Note that $G(\Omega_0=1, \Omega_\Lambda=0)=1$.

In the highly non-linear regime, the CF obtained from a scale-free primordial
spectrum obeys a simple scaling relation (Groth \& Peebles 1977).
One can interpolate between the linear and highly non-linear regimes to 
produce semi-analytic models for clustering in the quasi-linear
regime which can then be accurately fit to
results from $N$-body simulations (Hamilton 1991; Jain, Mo, \& White 1995; 
Peacock \& Dodds 1996). Following the discussion of PD, we define the 
dimensionless power spectrum as 
$\Delta^2(k,z)\equiv (2\pi^2)^{-1} k^3 P(k,z)$. The non-linear
dimensionless spectrum is then given by 
     $\Delta_{NL}^2(k_{NL},z) = f_{NL}[ \Delta^2(k,z)]$,
where the linear and nonlinear scales are related by
$k=[1+\Delta_{NL}^2(k_{NL})]^{-1/3} k_{NL}$.  The form of $f_{NL}$ 
for various values of $\Omega_0$ is given in PD.

\begin{figure}
\centerline{
\psfig{figure=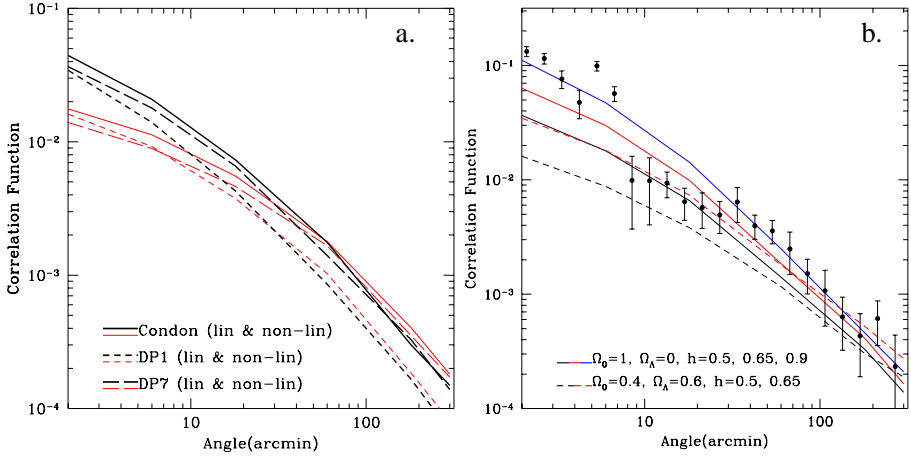}
}
\caption {CDM predictions for $w(\theta)$. (a) shows the effect of varying the assumed redshift distribution and the effect of including non-linear evolution in an $\Omega_0=1$, $\rm H_0=50$ universe (bold curves include non-linear evolution). (b) shows the effect of varying the cosmological parameters using the DP7 redshift model (CF amplitude increases with increasing $\rm H_0$).}
\end{figure}

Once we have an expression for the power spectrum (including
contributions from non-linear evolution) at all epochs, we
can Fourier transform this and substitute into Limber's equation
to obtain an estimate of the predicted angular CF.  

Figure 2(a) shows how the predictions for $w(\theta)$ vary for different 
estimates of the redshift distribution and how they change when
non-linear evolution is considered. There appears to be a fair amount
of uncertainty in the predicted result, depending on which $dN\over dz$
model is chosen.
It should be noted, however, that Condon's luminosity
function is more carefully constructed to fit the low-$z$ population
and it is these sources that one expects to be contributing
most to the clustering signal (see below). In addition, Peacock has 
indicated that DP's model 7 is the best model to use. If one limits
oneself to only these two models then uncertainties in the redshift
distribution do not contribute very significantly to uncertainties in
the predicitions. The effect of non-linear evolution is significant
on scales less than $\sim30^\prime$. 

%\subsection{Discussion}
\section{Discussion}
It is useful to investigate how one expects the contribution to the
clustering signal to vary with redshift. Since Condon's model and
DP7 include a large local population, these two models predict a
large contribution to the clustering signal from `local' sources:
$\sim60\%$ of the signal at $30^\prime$ is expected to come from sources 
with $z<0.1$ and this fraction increases as one goes to larger angles.
A scenario where the very nearby population is made up of a mix of
spirals and ellipticals (starbursting galaxies and AGN) and the
slightly more distant population begins to be dominated by AGN
is consistent with a clustering amplitude fairly close to that
of `normal' galaxies plus a slope more similar to that
observed for elliptical galaxies ($\gamma\sim 2.1$ for ellipticals).

Figure 2(b) shows the effect of varying cosmological parameters on the
predicted CF's but, given the above comments, it is somewhat misleading. 
It appears
that an $\Omega_0=1$ model with a high Hubble constant fits the
data best, but this assumes that all the sources
have similar clustering strengths. On large angular scales 
(larger than $\sim30^\prime$),
however, the CF is probing nearby structures and seems to be indicating a bias,
$b$, for these sources of a little more than 1 
(where $\xi_{radio}=b^2\xi_{mass}$).
On the smaller scales where one expects to be more sensitive to 
higher-$z$ clustering, a larger bias is required if models with popular values of $\rm H_0$ are to fit.
This is consistent with the difference in clustering strengths measured by
PN for local AGN ($10h^{-1}$Mpc) and that measured for optically observed 
galaxies. 
 
The angular CF of radio sources associated with $E>19$ galaxy
matches is significantly higher than what 
one would expect for a normal sample of $E>19$ APM galaxies. This
can be explained if the APM galaxies with radio counterparts are, 
on average, closer than all APM galaxies with $E>19$.
Ultimately, one would like to investigate
clustering at high-$z$, but this is difficult given that the angular CF
measured here appears rather sensitive to the clustering of a nearby 
population of sources which probably does not have the same clustering
strength as the population at higher-$z$. In a preliminary attempt to isolate
the clustering of the AGN-dominated sample at higher-$z$, we determined the CF
for all sources $without$ galaxy counterparts. The $r_o$'s inferred from this
measurement (with a linear evolution assumption) ranged from 
$10h^{-1}$Mpc to $15h^{-1}$Mpc depending on the
redshift model. If the redshift distribution were better determined
in the moderate-$z$ range (beyond what one expects for APM galaxies,
but not so far that the projected clustering becomes negligible),
we could compare the clustering in this range with PN's measurement 
for local AGN to investigate the evolution of clustering in a more homogenous
population. 

\section{Weak Lensing}
Since it is not clear how light traces mass, it is desirable to 
find methods of probing mass fluctuations directly. In optical images, 
the weak gravitational lensing (coherent distortion) of background 
sources by foreground matter distributions has been used to probe 
cluster-scale masses and the use of similar techniques to probe larger
mass scales has been discussed (see Kaiser 1996 and references therein).
Here, we outline a preliminary search for a weak lensing signal in FIRST. 
A detailed discussion will be given in Refregier et al. (1997).
 
%While weak lensing (in optical images) has been successfully
%used to probe mass fluctuations on cluster scales, probing 
%fluctuations on larger scales in the optical requires deep images on 
%larger areas of sky than those obtained until now.  
%Since the new radio surveys provide a large number of sources
%at high-z and cover a large region of sky, it seems worthwhile to 
%investigate the possibility that the coherent distortion of radio source 
%images by foreground 
%matter could be used as a probe of the matter distribution.

We start by defining the ellipticities of the sources. If $a$ and $b$
are the major and minor axes of a source and $\alpha$ is its position
angle measured relative to some arbitrary set of axes defined on the sky, 
then stretching along these axes can be measured by 
$\epsilon_+=\epsilon\cos(2\alpha)$
and stretching at $45^{\circ}$ to these axes is measured by 
$\epsilon_{\times}=\epsilon\cos(2\alpha)$ 
[where $\epsilon=(a^2-b^2)/(a^2+b^2)$]. It is 
helpful to define ellipticity correlation functions which are independent 
of the coordinate system. To do so, we define $\epsilon^r_+$ and 
$\epsilon^r_{\times}$ measured with respect to axes which are parallel
and perpendicular to the line connecting 2 points which are being 
correlated. If $\phi$ is the angle between the sky axes and the axes defined
by the two points then one gets
$\epsilon_+^r = \epsilon_+ \cos 2 \phi + \epsilon_\times
\sin 2\phi$ and $\epsilon_\times^r = -\epsilon_+ \sin 2 \phi + 
\epsilon_\times\cos 2\phi$.

One can then construct three ellipticity CF's,
$C_1(\theta) \equiv \VEV{\epsilon_+^r(\vectheta_0) \epsilon_+^r(\vectheta_0+\vectheta)}$, 
$C_2(\theta) \equiv \VEV{\epsilon_\times^r(\vectheta_0)
\epsilon_\times^r(\vectheta_0+\vectheta)}$ and 
$C_3(\theta) \equiv \VEV{\epsilon_+^r(\vectheta_0)\epsilon_\times^r(\vectheta_0+\vectheta)}$. Under reflection along the line connecting 2 points, 
$\epsilon^r_+$ is invariant but $\epsilon^r_\times$ changes sign, so
$C_3$ must always be zero. 

These three CF's have been measured for sources in
the FIRST catalog that are not `point' sources, i.e for all sources
that have $a>2^{\prime\prime}$. With this cut, there are about
60 sources $\rm deg^{-2}$. The ellipticities used were those obtained
$before$ beam deconvolution, making the measured CF's only rough
estimates of the true CF's. To obtain an estimate of the noise,
the CF's were also 
measured for a randomized catalog in which sources are assigned random
ellipticities. We found that the $C_1$'s and $C_2$'s measured in the catalog 
were well above the noise. The
$C_3$'s we measured were consistent with zero. While these results are
encouraging, systematic effects are likely to contribute significantly to the 
signal and we are still studying these in detail. 

We have also investigated what CDM models predict for $C_1$     
and $C_2$. A COBE normalised model with $\Omega=1$ and $h=0.5$ predicts
values which are above the noise. Normalising to popular values of
$\sigma_8$, however, decreases the predicted signal. Further work is
required before we will know if a weak lensing signal is detectable but if 
it were, it would provide a unique opportunity to probe mass fluctuations
on scales that are very difficult to probe with optical surveys. 

ACKNOWLEDGEMENTS
Collaborators include Marc Kamionkowski, David Helfand, Bob Becker,
Rick White, Michael Gregg, Alexandre Refregier, Scott Brown \& Richard McMahon.
The work was supported by grants from NASA (contract NAG5-3091), 
the National Geographic Society,
the NSF, NATO, IGPP, Columbia University and Sun Microsystems.

\end{document}